# Cross-Bridge Induced Adhesion of Red Blood Cells Assessed by Optical Tweezers

A. Popov*, T. Avsievich, A. Bykov and I. Meglinski*

*Optoelectronics and Measurement Techniques, University of Oulu, Oulu P.O. Box 4500 FI-90014, Finland*

*Correspondence: alexey.popov@oulu.fi and/or igor.meglinski@oulu.fi

The reversible aggregation of red blood cells (RBCs) is a process of erythrocytes clumping that strongly influences the rheological properties of blood. The adhesion of RBCs has been studied extensively in the frame of cell-to-cell interaction induced by dextran macromolecules, whereas the data is lacking for native plasma solution. We apply optical tweezers to investigate the induced adhesion of RBCs in plasma and in dextran solution. Two hypotheses 'cross-bridges' and 'depletion layer' are being utilized to describe the mechanism of cells interaction, while both need to be confirmed experimentally. The results show that in dextran solution the interaction of adhering RBCs agrees well with the quantitative predictions obtained on the basis of a depletion-induced cells adhesion model, whereas a migrating 'cross-bridge' model is more appropriate for plasma. Furthermore, the 'cross-bridge' mechanism is confirmed by direct visualization of red blood cells adhesion utilizing scanning electron microscopy (SEM).

The aggregation of red blood cells (RBCs) process is the reversible clumping of erythrocytes that can dramatically influence the rheological properties of blood. The increased aggregation of RBCs is a high risk factor known in a number of pathologies, including malaria, diabetes mellitus, hypertension, inflammation and others [1]. For decades, fundamental studies of RBCs aggregation have been attracting a significant attention from various viewpoints ranging from basic biophysical examination to perspectives of clinical application. However, the basic mechanisms of this process are still not thoroughly understood.

Various techniques have been utilized for quantitative characterization of RBCs aggregation. The dynamic light scattering was applied for assessment of aggregation of an ensemble of RBCs for diagnostic purposes [2]. Conventional microscopy, since its first application, is widely used to study the relation of RBCs aggregation to pathological conditions [3,4]. Micropipette aspiration technique (MAT) was successfully employed for the measurements of energy of cell-to-cell interaction [5]. Scanning electron microscopy (SEM) is routinely utilized for direct observation of cells interaction, their deformation and variations of intercellular distances [6]. Atomic force microscopy (AFM) is used effectively for quantitative assessment of the parameters of RBCs' adhesion [7].

Introduced by Ashkin, the trapping of micro-particles with a sharply focused laser beam [8], so-called optical tweezers (OT), is now becoming a popular tool, in particular for studying the interaction of cells [9,10]. OT allows to measure the forces of interaction between cells with up to pico-newton (pN) resolution, providing an opportunity for a precise control over cells contact [11]. This approach provides a high potential to gain new insights in the mechanisms of RBCs interaction. In the current report we apply the OT technique developed in-house [12,13] for the quantitative assessment of RBCs' adhesion, with the ultimate goal to confirm the hypothesis of 'cross-bridges' or 'depletion layer' induced adhesion of RBCs.

The aggregation of RBCs is influenced by both cell-specific and extracellular factors. In native solution (plasma) the latter are represented by a number of relatively big proteins such as fibrinogen, globulins, C-reactive proteins and others [1]. The solutions of neutral macromolecules, such as dextran, are used to induce RBCs aggregation that resembles the one in plasma [14]. Dextran solution serves as a probe for cell-specific parameters influencing RBCs aggregation.

An interaction between single RBCs has been analyzed based on the so-called 'depletion interaction' model [15]. In this model the interaction between cells is induced by macromolecule depletion near cells due to the balance of entropy and the resulting osmotic pressure pushing cells towards each other, thus forming aggregates [16]. The quantitative calculations of the interaction energies of adhering RBCs obtained in the 'depletion layer' model were found to perfectly match the results of direct measurements performed by the AFM and MAT [5,7].

An alternative model developed for qualitative description of RBCs interaction is based on the concept of the adsorption of macromolecules to membranes of adjacent cells and formation of 'cross-bridges' [17]. The key advantages of this model are the independence of the macromolecule concentration from the intercellular distances between cells and specific binding of macromolecules to RBCs membrane. The drawback of the model is that an electrostatic repulsion between RBCs makes it impossible for cells to approach each other close enough to start forming cross-bridges.

It should be pointed out that both models mentioned above were developed and tested for the dextran-induced aggregation of cells, while there is a very limited data available for the native plasma solution [16].

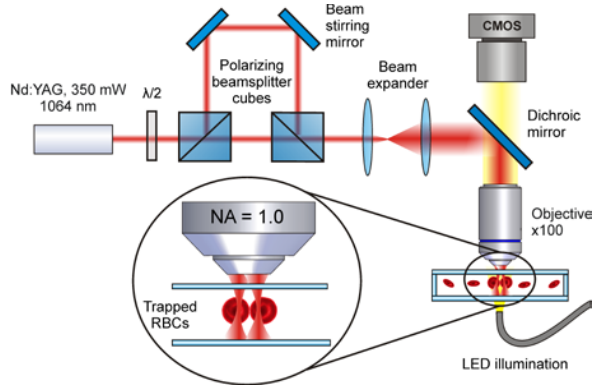

FIG.1. Schematic presentation of the two-channel OT setup for measurements of RBCs interaction. Optical traps were formed with a water-immersion objective (×100, NA = 1.00). Two interacting RBCs were stretched with two OTs at the edges. The image is formed in transmission mode and registered using a CMOS camera (Pixelink PL-B621M, Canada).

An OT experimental system developed in house [12,13] was applied for the study of RBCs interaction (Fig.1) both in dextran and native plasma solutions. Two optical traps were formed by orthogonally-polarized continuous-wave laser beams from a single-mode Nd:YAG infrared laser ($\lambda$ = 1064 nm, 350 mW, ILML3IF-300 Leadlight Technology, Taiwan) and a large numerical aperture (NA = 1.00) 100× water-immersion objective (Olympus, LumPlanFI, USA). Laser power in each trap varied in the range between 3 and 30 mW. Cells heating was negligible at the given laser power. The position of traps was controlled within the objective focal plane by a beam stirring mirror with a step-motor translation stage (Standa). The setup allows to move the trap position in a range of velocities ~ 0.1-100 µm/sec. Visual control of the trapped objects was done in a transmission configuration using the CMOS camera. Two individual RBCs are trapped to measure the interaction energy by OT. The measurements were carried out on 100 pairs of RBCs.

Fresh RBC samples were obtained from a healthy donor by finger prick. The blood sample was placed in phosphate buffer saline (PBS) and washed by centrifugation for 10 minutes at 6500 RPM (4000g). The dipotassium ethylenediaminetetraacetic acid (K2-EDTA) was used as an anticoagulating agent. The experiments were performed at the room temperature within 4 hours after drawing blood. The measurements were performed with the RBCs suspended in (1) platelet-free plasma and (2) PBS (Invitrogen, USA) solution of 5 mg/ml dextran 500 kDa (Pharma, USA). To separate plasma from RBCs, centrifugation of 4 ml of blood at 1800g was performed for 10 minutes. Second centrifugation at 12000 g for 10 minutes was implemented to remove the remaining platelets. Washed RBCs were suspended in the platelets-free plasma and in the dextran solution with the ratio of 0.05%. A chamber used for measurements consisted of two glass plates separated by a 100 µm gap made with adhesive tape. About 60 µl of the solution were added to the chamber. For experiments with dextran solution, glass surfaces were coated with 5 µl of 1% PBS solution of human serum albumin (Sigma-Aldrich, USA) and dried to prevent morphological changes of RBCs due to their interaction with the glass surface.

The flow is created by moving the sample of suspended RBCs with a motorized 2D stage, and the velocity of the flow is step-wisely increased until the trapped cell escapes from the trap. The trapping beam power is being kept constant, and when the cell is escaping the trap the force is considered to match the trapping force. Thus, the cells interaction is measured by matching it with the escape force.

The results of measurements of the RBC adhesion by OT are presented in Fig.2.

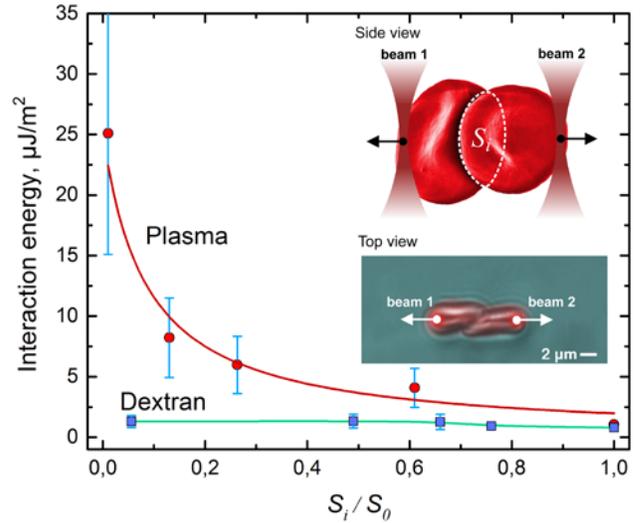

FIG.2. Interaction energy of the adhering RBCs as a function of relative conjugated surface area to the total initially overlapping area derived from the measurements in: the dextran 500 kDa (5 mg/ml) solution (squares) and plasma (circles). Solid lines are predictions based on the migrating 'cross-bridges' model for plasma (red), and the 'depletion interaction' model for dextran solution (green). The interaction area $S_i$ between the cells is counted based on the direct measurements of cells relative displacement and their size.

As one can see, in the dextran solution the interaction energy of adhering RBCs is found to be unchanged during the stretching of the cells by OT (see Fig.2). Based on the 'depletion interaction' model [15], this is explained by the fact that osmotic forces are proportional to the interaction area and uniformly distributed across the cells surface. The 'depletion layer' model combines electrostatic repulsion caused by RBC surface charge and osmotic attractive forces generated by polymer depletion near the RBC surface and penetration of macromolecules into glycocalyx.

Following the 'depletion interaction' model, the interaction energy of adhering RBCs is defined as [15]:

$$E = -2p(\Delta - \frac{d}{2} + \sigma - z) \quad (1)$$

Here, $p$ is the osmotic pressure, $\Delta$ is thickness of the depletion layer, $d$ is the separation distance between adjacent surfaces, $\sigma$ is the glycocalyx thickness, and $z$ is the depth of polymer penetration into the glycocalyx. We found that the mean value of the interaction energy $<E_{dextran}> = 1.2 \pm 0.6$ µJ/m$^2$, observed experimentally by utilizing the OT approach (see Fig.2), agrees well with the value predicted by the 'depletion interaction' model (~1 µJ/m$^2$) [15].

When the RBCs are suspended in plasma the interaction energy grows significantly with the tension of cells by OT and the 'depletion layer' model becomes invalid. In fact, the obtained results are in excellent agreement with the predictions based on the migrating 'cross-bridges' model (see Fig.2).

In the migrating 'cross-bridges' model the interaction energy ($E$, [µJ/m$^2$]) is defined as [18]:

$$E = 2kTm_0 b \left[1 + b\left(\frac{S_i}{S_0}\right)\right]^{-1} \quad (2)$$

Here, $S_i$ is the conjugated surface area between two interacting cells, $S_0$ is the initial interaction (overlapping) area ($S_0 = 25$ µm$^2$), $b$ is the dimensionless binding affinity, $m_0$ is the initial cross-bridge density before adhesion (1/nm$^2$), $k$ is the Boltzmann constant, $T$ is the absolute temperature ($kT = 4 \times 10^{-21}$ J).

Thus, based on the results of OT measurements (see Fig.2) we found values of the binding affinity coefficient $b = 11 \pm 5$ and the cross-bridge density $m_0 = 1/3600$ nm$^{-2}$.

In addition, Sigma field emission SEM (Carl Zeiss, Germany) were applied to visualize the process of the RBCs adhering. The results of a direct visualization of the RBCs just before their separation is shown in Fig.3. The obtained images clearly show the discrete cross-bridges at the surface of RBC membrane. As one can see, the cross-bridges are identical and uniformly distributed across the interaction areas (see Fig.3). Density of cross-bridges obtained by SEM imaging (1-2 per 100 nm distance, see Fig.3) agrees well with the value predicted by the migrating 'cross-bridges' model ($m_0 \sim 1.6$ per 100 nm). Actual size of a 'cross-bridge' was estimated as ~ 30 – 40 nm.

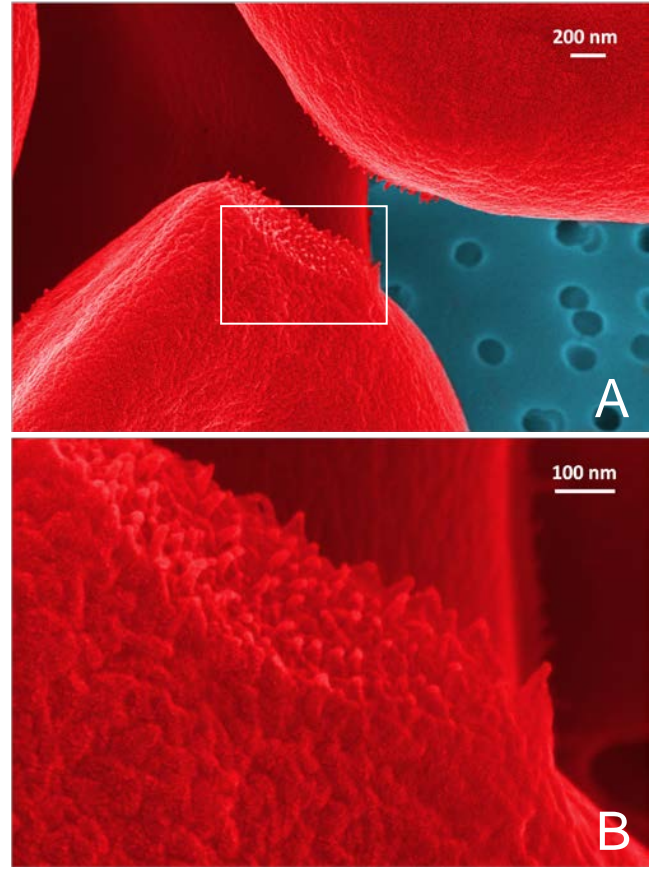

FIG.3. Colour-coded SEM images of the separated RBCs (A). White rectangle indicates an enlarged area on the cell surface with cross-bridges, separately shown in (B).

In summary, we have demonstrated that OT provides a unique opportunity to obtain new insights into mechanisms of red blood cells aggregation, and can be used as an effective tool for quantitative assessment of the RBCs interaction. We have applied the OT technique to investigate interaction of RBCs in dextran solution and in plasma. The OT approach has been used to measure the interaction energy of the adhering RBCs as a function of the relative conjugated surface area. We found that in the dextran solution the interaction of adhering RBCs agrees well with the quantitative predictions obtained on the basis of the depletion-induced cells adhesion model. We also found that the migrating 'cross-bridge' model is more appropriate for the quantitative description of interaction of the adhering RBCs in plasma. This model describes the accumulation of interaction energy due to formation of 'cross-bridges' on adjacent cell membranes. By utilizing SEM, this mechanism has been confirmed by direct visualization of 'cross-bridges' at the surface of RBC membrane, for the first time to our knowledge. The density of cross-bridges, assessed by the OT measurements, utilizing the migrating 'cross-bridge' model, is in excellent agreement with the results of SEM visualization.

The authors acknowledge financial support of Academy of Finland (projects 260321, 290596) and CIMO Fellowship (grant TM-17-10370). IM acknowledges partial support provided by RSF (project №15-14-10008). Authors acknowledge initial involvement of Dr. K. Lee and Dr. A.V. Priezzhev from Moscow State University (Russia) and Dr. A. Karmenyan from the Institute of Biophotonics of National Yang-Ming University (Taipei, Taiwan) at the early stage of the OT experiment development.